\documentclass{nature}
\usepackage{graphicx}
\makeatletter
\let\saved@includegraphics\includegraphics
\AtBeginDocument{\let\includegraphics\saved@includegraphics}
\renewenvironment*{figure}{\@float{figure}}{\end@float}
\makeatother

\usepackage{xfrac}
\usepackage{multirow}
\usepackage{subfigure}
\usepackage{mathtools}
\usepackage[colorlinks=true,linkcolor=blue,citecolor=black,urlcolor=blue]{hyperref}
\usepackage{authblk}
\usepackage{lineno}
\usepackage[labelfont=bf]{caption}
\usepackage[figurename=Figure ]{caption}
\usepackage{siunitx}

\begin{document}

\title{Sub-nanoscale atom-by-atom crafting of skyrmion-defect interaction profiles}
\author{I Gede Arjana$^{\text{1}}$, Imara Lima Fernandes$^{\text{1,*}}$, Jonathan Chico$^{\text{1}}$ \& Samir Lounis$^{\text{1,*}}$}
%\affil[1]{Peter Gr\"{u}nberg Institut and Institute for Advanced Simulation, Forschungszentrum J\"{u}lich and JARA, D-52425 J\"{u}lich, Germany}
%\affil[*]{i.lima.fernandes@fz-juelich.de, s.lounis@fz-juelich.de}

\maketitle

\begin{affiliations}
	\item[$^{\text{1}}$ ] Peter Gr\"{u}nberg Institut and Institute for Advanced Simulation, Forschungszentrum J\"{u}lich and JARA, D-52425 J\"{u}lich, Germany
	\item[$^{\text{*}}$ ] i.lima.fernandes@fz-juelich.de, s.lounis@fz-juelich.de
\end{affiliations}

\begin{abstract}
Magnetic skyrmions are prime candidates as information carriers for spintronic devices due to their topological nature and nanometric size.  However, unavoidable inhomogeneities inherent to any material leads to pinning or repulsion of skyrmions that, in analogy to biology concepts,  define the phenotype of the skyrmion-defect interaction, generating complexity in their motion and challenging their application as future bits of information. Here, we demonstrate that atom-by-atom manufacturing of multi-atomic defects, being antiferromagnetic or ferromagnetic, permits the breeding of their energy profiles, for which we build schematically a Punnet-square. As established from first-principles for skyrmions generated in PdFe bilayer on Ir(111) surface, the resulting interaction phenotype is rich.  It can be opposite to the original one  and eventually be of dual pinning-repulsive nature yielding energy landscapes hosting multi-domains. This is dictated by the stacking site, geometry, size and chemical nature of the adsorbed defects, which control the involved magnetic interactions.  This work provides new insights towards the development of disruptive device architectures incorporating defects into their design aiming to control and guide skyrmions.
\end{abstract}

Magnetic skyrmions~\cite{Bogdanov,Roessler2006}, i.e. non-collinear spin textures with particle-like properties, are promising future magnetic bits for future data storage technologies based on topological concepts~\cite{Fert2013,Sampaio2013,Tomasello2013,Zhou2014,Crum2015,Zhang2015,Yu2016,Garcia-Sanchez2016,Xia2017}. Of great technological relevance are skyrmions in thin films and magnetic multilayers~\cite{Heinze2011,Romming2013,Jiang2015,Woo2016,Boulle2016,Wiesendanger2016,Maccariello2018,Meyer2019,Zhang2020,Duong2019,Kang2018}, which can be stabilized as a result of the competition among the Heisenberg exchange interaction (HEI), Dzyaloshinskii-Moriya interaction (DMI)~\cite{Dzyalosinkii,Moriya} and the perpendicular magnetic anisotropy. The low spin polarized current thresholds required to manipulate skyrmions compared  to typical ferromagnetic domain walls~\cite{Yu2012,Fert2013} along with their high mobility, high stability and small sizes make them ideal for achieving efficient and functional devices. However, defects that are ineluctable in any device and materials are often seen as inhibitors for applications.

The dynamical behavior of the skyrmion motion as function of applied currents hinges on the presence of defects, which define the three motion regimes: pinning, creep- and steady-flow-motion as demonstrated experimentally in ultrathin heavy metal/ferromagnetic bilayers and multilayers~\cite{Woo2016,Jiang2016,Litzius2017}.  Atomic-scale imaging based on scanning tunneling microscopy demonstrated that skyrmions in PdFe bilayer on Ir(111) are inert to the presence of a single Co adatom, in accordance to recent ab-initio simulations~\cite{Fernandes2018}, but react to the presence of a Co trimer~\cite{Hanneken2016}.  Not only skyrmions experience pinning but also magnetic vortices with cores containing thousands of atoms. This was visualized with  
spin-polarized scanning tunneling microscopy, which was utilized to extract the pinning strength by moving vortices across defects with a vector magnetic field~\cite{Holl2020}.

In tandem to the few available experiments,  scarce ab-initio simulations addressed the case of point-defects~\cite{Choi2016,Fernandes2018,Fernandes2019} while several phenomenological-based  studies were dedicated to various defects assuming local changes in the magnetic properties of the material~\cite{Fert2013,Liu2013,Iwasaki2013a,Iwasaki2013b,Fert2017,Navau2018,Castell2019,Diaz2018,Stosic2017,Reichhardt2018,Brown2019,Brown2019b,Muller2017}. Yet, the energy-landscape impacting skyrmions as induced by complex defects remains weakly explored.

In the current work, we demonstrate that multi-atomic defects can behave in stark contrast to their single-atom counterpart when interacting with a magnetic skyrmion. 
The mechanisms favoring either skyrmion pinning or repulsion define two interaction genotypes. Similarly to concepts known in biology, they coexist in each defect but the observable interaction phenotype is one or the other depending on their relative strength. 
%The skyrmion-defect interaction results from the interplay of mechanisms, favoring either repulsion or pinning, defining the interaction genotype carried by the individual atomic defects.
The breeding of the  interaction profiles of the impurities by building-up multi-atomic defects may yield unexpected energy landscapes that can be engineered via various external means.  Not only the chemical nature of the inhomogeneities was found of paramount importance but also their shapes, size and stacking sites. This gives a plethora of opportunities for new device architectures incorporating defects into their design by manipulating either their attractive or repulsive phenotypes.

\begin{figure}
	\centering
	\includegraphics[width=\textwidth]{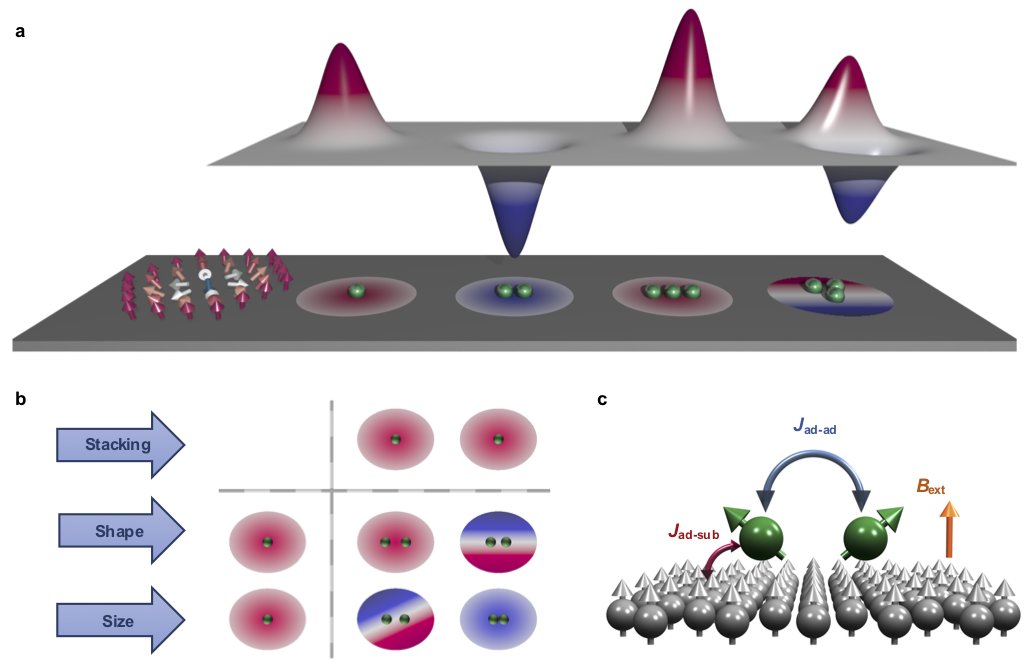}
	\caption{\textbf{Atom-by-atom manufacture of skyrmion-defects energy landscapes.} \textbf{a} A schematic picture of a magnetic skyrmion interacting with different types of defects indicated by green spheres. A plethora of interactions profiles emerge from building-up nanostructures made of the same atoms.   \textbf{b} Graphical representation of a Punnett square describing the possible nature of the skyrmion-defect interaction that arises from bringing two repulsive adatoms together. The isolated adatoms act symbolically as parents and ``carry'' a repulsive interaction phenotype (red color) resulting from the subtle competition of mechanisms defining pinning and repulsion genotypes. %while the type of interaction profile represents the investigated genotype.
	Owing to various mechanisms rooting in the magnetic interactions depicted in \textbf{c}, the resulting interaction phenotype can be trivially similar or even opposite to the one of the parents, and can remarkably develop a dual behavior.  }
	\label{fig:Pict2}
\end{figure}

\begin{figure}
	\centering
	\includegraphics[width=\textwidth]{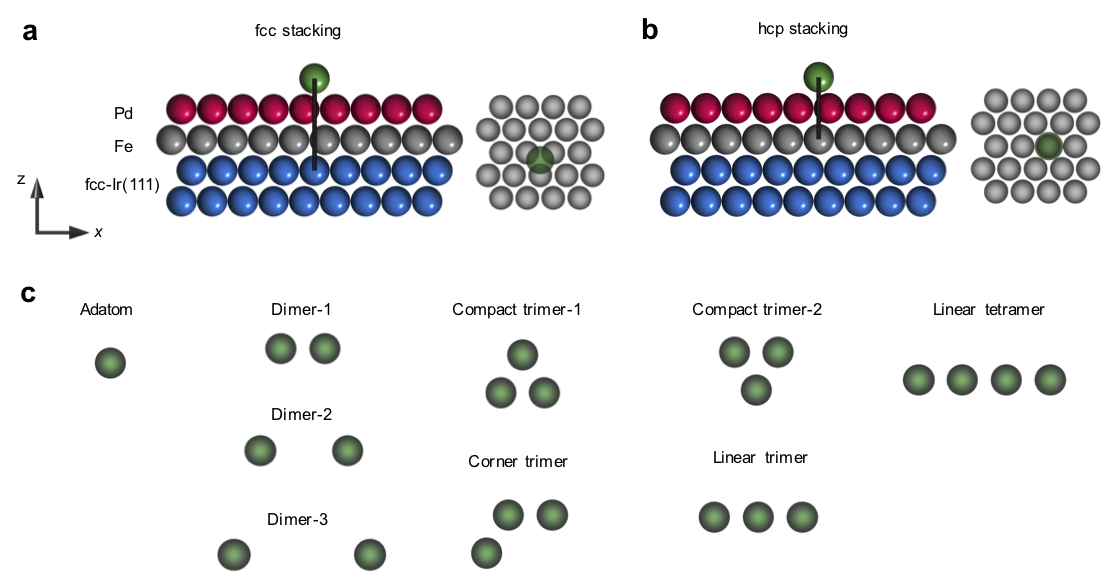}
	\caption{\textbf{Sketched representation of the investigated  nanostructures.} The adatoms can either sit on the \textbf{a} fcc or \textbf{b} hcp stacking configurations on fcc PdFe bilayer on Ir(111) with an adatom (green sphere) deposited atop the Pd layer. Red, grey, and blue spheres represent the Pd, Fe, and Ir layer, respectively. \textbf{c}  Various potential  geometrical configurations of the defects. }
	\label{fig:Pict3}
\end{figure}

We use a multiscale modelling approach, with parameters defining an atomistic extended Heisenberg model mapped from first-principles calculations (see Methods section) to investigate the energy profile of a single skyrmion  at the vicinity of defects (dimer, trimers and tetramers) of different shapes, sizes and chemical nature deposited on PdFe/Ir(111) surface (Figure~\ref{fig:Pict2}a).  The latter substrate is known to host few nanometers-wide magnetic skyrmions~\cite{Romming2013,Romming2015,Dupe2014,Simon2014,Dias2016,Leonov2016,Bouhassoune2019}. 
The defects consisting of 3$d$ transition metal atoms are located on top of the Pd layer and we consider both fcc and hcp stacking sites (Figure~\ref{fig:Pict3}a-b). The nanostructures can be ferromagnetic, antiferromagnetic, ferrimagnetic or even non-collinear when deposited on the saturated substrate. Here, we focus on nanostructures made of elements leading to unexpected interaction patterns such as Fe and Cr adatoms and illustrate their behavior with a collection of interesting examples. Although, single Fe and Cr adatoms tend to repel the investigated skyrmion, surprising new behavior emerges from placing the atoms in a cluster as shown schematically  in   Figure~\ref{fig:Pict2}b in the form of a Punnett square. While the trivial outcome of breeding two identical interaction profiles when forming a dimer is to generate the same phenotype profile, the most astonishing result is the emergence of an interaction profile of opposite behavior with respect to the phenotype of the isolated adatoms that act as ``parents''. Even more remarkable is the appearance of a dual behavior exhibiting both pinning and repulsive regions. The competition between the intra-defect and defect-substrate magnetic interactions  are found to play a substantial role in establishing these counter-intuitive behaviors (Figure~\ref{fig:Pict2}c).

\section*{Results}

\subsection{Overview of the nature of skyrmion-defects interactions.}

The interaction energy between the various defects and the single magnetic skyrmion is defined as the energy difference between the two cases: (a) the skyrmion at the vicinity of the defect and (b) the skyrmion and defect far away from each other. A negative (positive) energy difference indicates an attractive (repulsive) skyrmion-defect interaction.

Our study is based on the following atomistic extended Heisenberg hamiltonian with parameters extracted from first-principles calculations (see Methods section for details):
\begin{equation}
     \mathcal{H}= - \frac{1}{2} \sum_{i \ne j} J_{ij} \hat{m}_i \cdot \hat{m}_j - \frac{1}{2} \sum_{i \ne j}^{}  \vec{D}_{ij} \cdot \left(\hat{m}_i \times \hat{m}_j\right) -  \sum_{i}^{} K_i\left(\hat{e}^z \cdot \hat{m}_i\right)^2 - \sum_{i}^{} \vec{M}_i \cdot \vec{B}_\text{ext},
\label{Eq.JM_interaction2}
\end{equation} 

\noindent  where the unit vector $\hat{m}_{i} = \vec{M}_{i}/M_i$ defines the direction of the atomic magnetic moment $\vec{M}_{i}$ at site $i$. The first term, $J$, describes the HEI, the second is the DMI, $\vec{D}$, while $K$, corresponds to the uniaxial magnetocrystalline anisotropy energy (MAE) being positive for an out-of-plane easy axis. The defects can have a MAE of different sign and magnitude than that of the substrate (see Supplementary Table 1). This, however, does not affect the orientation of the moments which is imposed by the adatom-substrate HEI (see Table~\ref{tab:jij}). The last term is the Zeeman contribution due to an external magnetic field $\vec{B}_\text{ext}$ applied along the magnetization of the ferromagnetic substrate, i.e. the direction perpendicular to the substrate, which defines the z-direction. A field of 10 Tesla is considered in this study, which stabilizes single magnetic skyrmions with a diameter of about 3.2 nm.%, defined as $\theta_{\text{R}_{\text{sk}}} = 90^\circ$. 

The adatom-substrate magnetic interactions oscillate as function of distance with the nearest neighboring (NN) one being the most relevant (see Supplementary Note 1 and Supplementary Figure 1). Considering the NN-averaged HEI presented in Table~\ref{tab:jij}, we expect the Fe nanostructures to be ferromagnetic by virtue of the first Hund's rule.  In contrast, the antiferromagnetic interactions of the Cr nanostructures (see Table~\ref{tab:jij}) lead to a rich set of magnetic textures (see for example Supplementary  Figure 2 obtained when the defects are away from the skyrmion). 

\begin{figure}
	\centering
	\includegraphics[width=\textwidth]{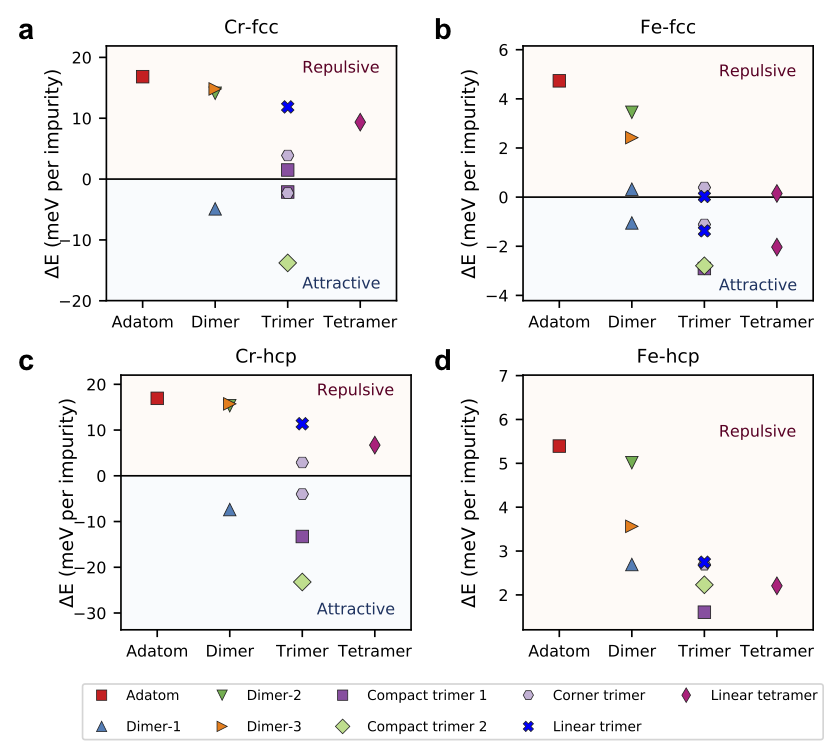}
	\caption{\textbf{Extrema of the skyrmion-defect binding energies.} Positive (negative) energies indicate repulsion (pinning) of a magnetic skyrmion by \textbf{a},\textbf{c} Cr and \textbf{b},\textbf{d} Fe nanostructures. Top and bottom panels present the results obtained for the fcc and hcp stacking sites, respectively. The size, shape and stacking have a tremendous impact on the  binding energies: some of the defects switch their interaction nature from repulsion to pinning and vice-versa and eventually host interactions of opposite sign. When the latter happens, two symbols assigned to the same nanostructure.   }
	\label{fig:Co}
\end{figure}

After calculating the energy profile for all the investigated nanostructures depicted in Fi\-gure~\ref{fig:Pict3}c, we collect systematically the maximum (minimum) repulsive (attractive) skyrmion-defect binding energy per impurity as a function of the cluster size in Fi\-gure~\ref{fig:Co}.
The breath of the binding energies, i.e. largest positive and negative values, is distinctly larger (a factor of three) for Cr-clusters as compared to that of Fe-clusters. This is related, in particular, to the adatom-substrate HEI being the strongest for Cr (see Table~\ref{tab:jij}).
 Separating the adatoms, as done for the dimers depicted in Figure~\ref{fig:Pict3}c, weakens the hybridization channels among their electronic states, which drastically reduces the adatom-adatom magnetic interactions shown in Table~\ref{tab:jij} and allows to recover the magnetic and interaction behaviors of the individual adatoms.

One notices that while single Cr-adatoms are repulsive, Cr-dimers can be pinning if the two atoms are nearest neighbors (dimer-1). Thus, while the interaction-genotype was initially repulsive, it can switch to pinning if the adatom-adatom interaction is permitted. Even more notable is that adding other Cr-adatoms to form a linear trimer or a tetramer, the resulting interaction phenotype switches back to the original repulsive nature. Intriguingly, the stacking site seems to mainly affect compact trimer-1 binding energies. 
 
For several cases, one notices the dual behavior of the binding energies. This means that the same nanostructure on the same stacking site  generates remarkably domains with opposite interaction-type with the skyrmion, i.e. repulsive and pinning regions surrounding the defect. In that case two points are assigned to the same defect. As displayed in Figure~\ref{fig:Co}, the stacking site has a more dramatic impact on Fe- than on Cr-nanostructures, which can be assigned to the weaker adatom-substrate HEI of the former compared to that of the latter. While all hcp Fe-clusters repel skyrmions, fcc ones experience a transition from repulsion towards either pinning or  a dual behavior when the size of the defects is increased.

As elaborated in the following subsections, the mechanisms driving the skyrmion-defect interaction genotype is settled by the chemical nature of the defects, their geometry, size and stacking sizes, which tune the balance between competing mechanisms affecting: (i) the magnetic interaction among the adatoms, (ii) the magnetic interaction of the adatoms with the Fe substrate, (iii) and  
Zeeman energy.  On the one hand, the 
defects tend to decrease the HEI among the neighboring Fe atoms of the substrate, initially reaching a value of 19.8 meV, owing to the hybridization of various electronic states (see Table~\ref{tab:jij}). 
 This  favors non-collinearity among the substrate spin moments, which tends to stabilize skyrmions at the vicinity of the defects. On the other hand, the HEI between the defects and the Fe substrate  provides an additional magnetic exchange interaction which tends to stiffen the surface magnetization disfavoring the presence of a magnetic skyrmion.  
 These mechanisms affect in a subtle fashion the HEI- and DMI-contributions, which usually counteract each other, to the skyrmion-defect binding energies (see also Supplementary Figure 2). Furthermore, Zeeman and magnetic anisotropy energies can favor pinning or repulsion if the difference between  the HEI- and DMI-contributions is not large enough. 
 
 Notably, reshaping and increasing the size of the nanostructures from dimers, to trimers (being of compact, corner and line forms) and tetramer open hopping channels for the electrons participating in the hybridization mechanisms within the nanostructure and with the substrate. This controls the magnitude of the magnetic interactions between the adatoms~\cite{Mavropoulos2010}, which is found to be in general the largest for the compact trimers (see Table~~\ref{tab:jij}), and dictates the overall magnetic behavior of the nanostructures. For conciseness, we focus our following analysis on the fcc stacking of the defects, but specific cases of the hcp stacking will also be addressed.

\subsection{Anatomy of defect-skyrmion interaction profiles: case of Cr-nanostructures.}

Single Cr adatoms display a large repulsive impurity-skyrmion interaction (see red square in Figure~\ref{fig:Co}) owing to electronic-structure-based mechanisms~\cite{Fernandes2018}, which can be simplified  resting on the aformentioned arguments. The adatom-substrate HEI, being more important than the defect-induced reduction of the substrate HEI that faciliates pinning, leads to a local stiffness of the spin-texture on the substrate, and its consequently repulsive behavior. As shown in Supplementary Figure 3 and Supplementary Note 2, the subtle balance between the DMI, HEI and Zeeman energies plays a significant role in defining the interaction phenotype of the adatoms. In fact, the DMI contribution to the binding energy counteracts the HEI contribution by about a factor of two.

\begin{figure}
	\centering
	\includegraphics[width=\columnwidth,keepaspectratio]{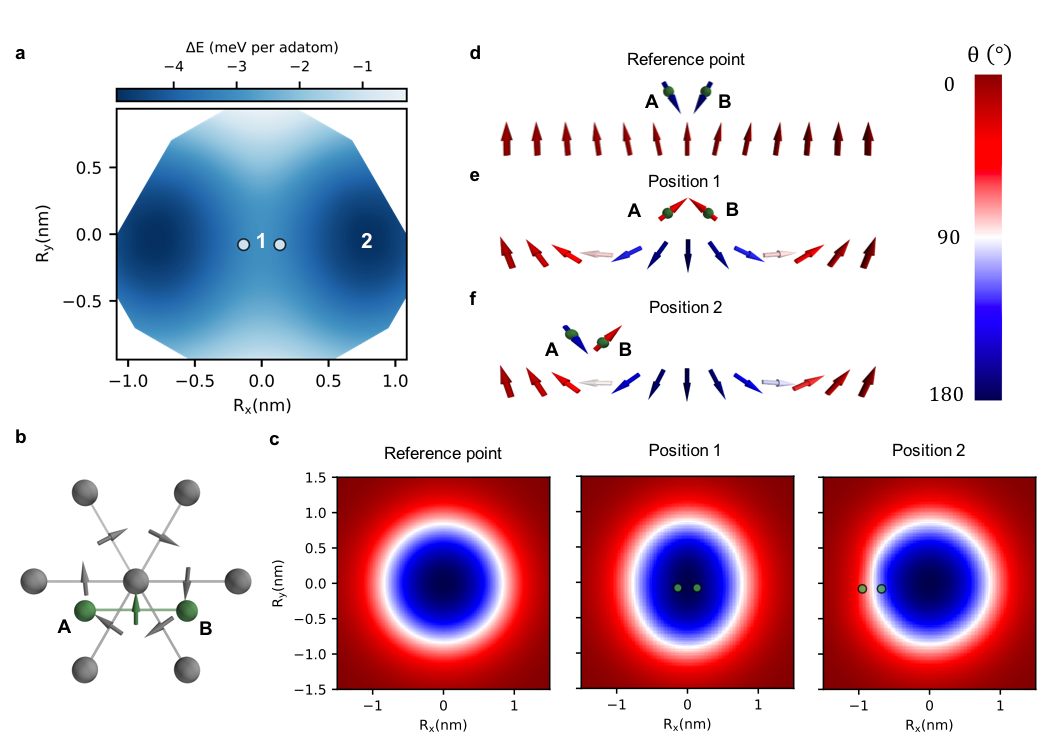}
	\caption{ \textbf{Energy profile and spin-texture of the skyrmion-dimer complex induced by fcc Cr dimer-1.} \textbf{a} The energy landscape around the dimer indicates a pinning behavior with  two minima at position 2. \textbf{b} DM vectors between a central Fe atom from the substrate and its nearest Fe neighbors (grey arrows) and the DM vector between the adatom A and the adatom B (green arrow). \textbf{c} The polar angle, $\theta$, of the skyrmion magnetic moments are plotted when the dimer is far away (reference point), at positions 1 and 2. The related magnetic states are depicted in the side views shown in  \textbf{d}, \textbf{e} and \textbf{f}, which illustrate how the magnetic moments of the dimer adapt to the skyrmionic spin-texture. At the vicinity of the skyrmion, the spin-chirality of the dimer switches as imposed by the skyrmion.  }
	\label{fig:CrDimer}
\end{figure}

Unexpectedly though, by forming the fcc Cr dimer-1, i.e. with the adatoms being first nearest neighbors, the skyrmion-defect interaction becomes attractive with a rather complex energy profile exhibiting two off centered symmetric minima, as depicted in Figure~\ref{fig:CrDimer}a. Here, magnetic frustration stabilizes the previously discussed spin-flop state similar to what was found on ferromagnetic substrates~\cite{Lounis2005,Lounis_Review_Wulfhekel,Lounis2007,Lounis2014}(see Supplementary Note 3). The  adatom-adatom and adatom-substrate HEI are simultaneously antiferromagnetic, which lead to a magnetic compromise where the Cr moments are antiferromagnetically aligned with a strong tilt opposite to the surface magnetization. When away from the skyrmion, the polar angle characterizing the adatom moments is $\theta = 165^\circ$ yielding an angle of $30^\circ$ between the two moments, with a left-handed chirality dictated by the intra-dimer DM vector. This result is inline with the angle of about $120^\circ$ obtained from the simplified Heisenberg model addressed in Supplementary Note 3  based on the averaged HEI given in Table~\ref{tab:jij} ($\cos{\theta} \approx - \frac{3}{2} \frac{\left<\text{J}_\text{ad-sub}\right>}{\left<\text{J}_\text{ad-ad}\right>}$).

By scrutinizing the spin-texture anatomy of the defect-skyrmion complex, we found that the skyrmion switches the spin-chirality of the dimer. When the skyrmion and the defects are far away from each other, and as previously discussed, the competition between the Cr-Cr and Cr-Fe HEI leads to a  non-collinear ground-state of the dimer with the left-handed chirality fixed by the DM vector between the adatoms (see green arrow in Figure~\ref{fig:CrDimer}b). However, the DM vector between the Fe substrate atoms has the opposite chirality (opposite DM interaction) as indicated by the gray arrows in Figure~\ref{fig:CrDimer}b, resulting in the formation of the Néel-type skyrmions with a right-handed chirality. Naturally, once the skyrmion is at the vicinity of the dimer, the competition between their opposite chiralities results in the chirality-switch of the dimer. The latter requires a lower energy cost than a local breaking and switching of the skyrmion's chirality so that the dimer spin texture follows the one of the underlying skyrmion.

The antiferromagnetic coupling between the Cr atoms leads to an almost  cancellation of the adatom-substrate exchange energy. Hence the reduction of the HEI among the substrate's Fe atoms prevails, which favours pinning. It should be noted that the minimum of the energy profile occurs when the fcc Cr dimer-1 is located on top of the region where the local magnetization of the skyrmion is mostly in-plane, i.e. close to the edge of the skyrmion as depicted by the  position 2 in Figures~\ref{fig:CrDimer}a-c. 

At  position 1, the polar angle of the adatoms moments is $51^{\circ}$ yielding to an opening angle of  $102^{\circ}$ between the two moments (see Figure~\ref{fig:CrDimer}d). At position 2 the polar angles are $140^{\circ}$ for atom A and $44^{\circ}$ for atom B with a resulting opening angle of $96^{\circ}$. Compared to the case where the skyrmion is far away from the dimer, the opening angle is larger at the skyrmion vicinity owing to the non-collinearity of the skyrmion weakening the effective adatom-substrate HEI. Therefore, the effective exchange energy provided by the dimer is diminished at the vicinity of the skyrmion,  which helps explaining the observed pinning. Likewise, the skyrmion is strongly reshaped when the cluster is on top of the skyrmion core as observed  in Figure~\ref{fig:CrDimer}c.  Accordingly, the canting between adjacent moments at the center of the magnetic skyrmion, position 1, decreases with respect to the shape obtained in the defect-free region. 
In contrast, the skyrmion recovers a more symmetric shape when its core is located at position 2. Thus,  the skyrmion is stiffer at position 1 than at  2, which explains the off-centered energy minima. In fact, dissecting the various contributions to the binding energies, we find that a major difference between the two positions lies in the DMI contribution being more repulsive for position 1 than for position 2 due to the observed stiffness (see Supplementary Figure 4).

%In fact, the axis, defined by the substrate magnetization, with respect to which the spin-flop state rotates across the skyrmion, dictating the Cr moments to rotate accordingly. 

\begin{figure}
	\centering
	\includegraphics[width=\columnwidth,keepaspectratio]{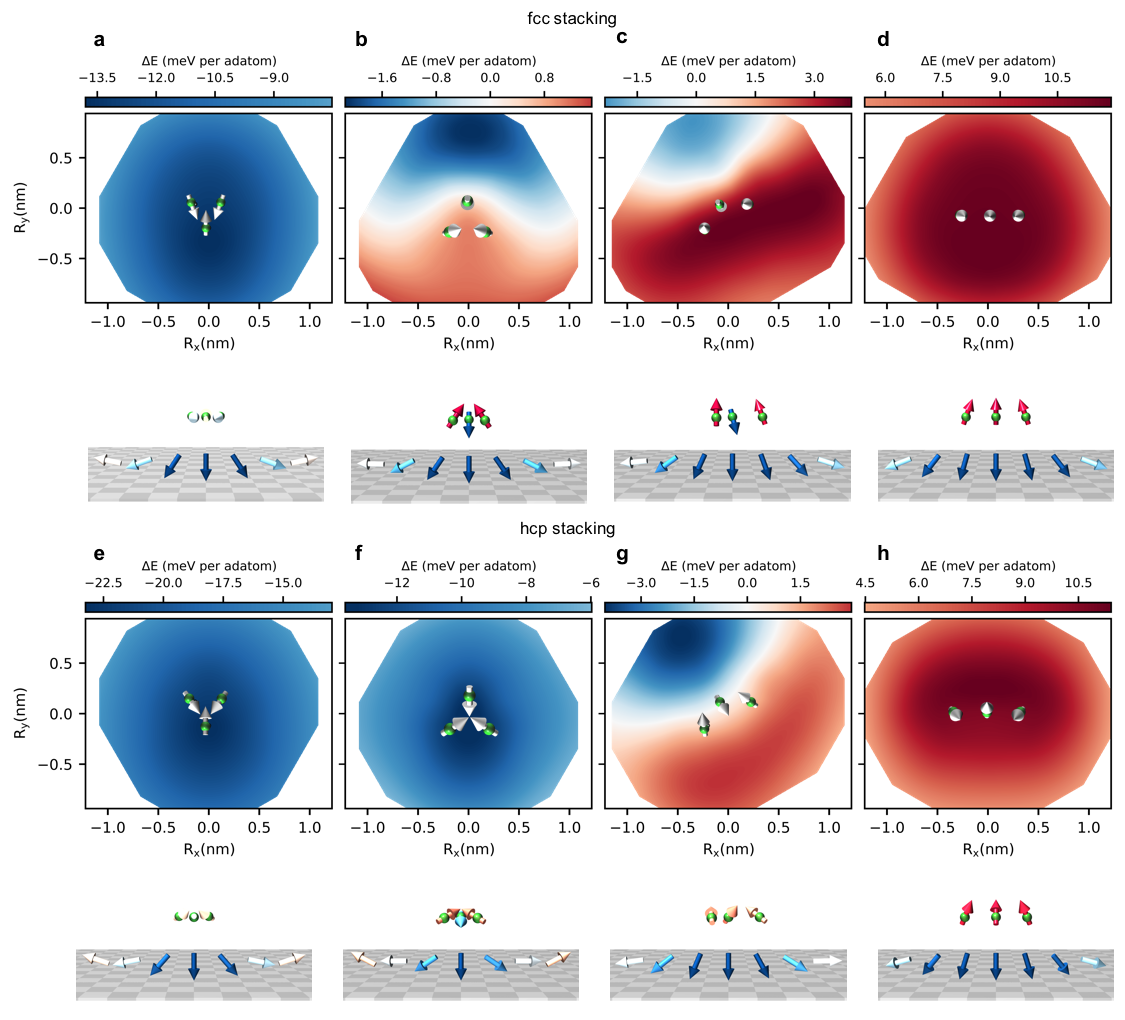}
	\caption{\textbf{Engineering of the skyrmion energy landscape by atomic manipulation of Cr trimers.} Surprisingly, repulsive domains emerge and grow in size after rotating an initially pinning compact Cr trimer-2, \textbf{a, e}, to obtain trimer-1, \textbf{b, f} and opening it to form a corner, \textbf{c, g}, and  linear trimers \textbf{d, h}. Upper (lower) panel corresponds to the fcc (hcp) stacking site. The magnetic texture of the trimers (green spheres) are illustrated when the skymrion's core is at position (0,0).   }
	\label{fig:Cr_trimer_fcc_hcp}
\end{figure}

Similarly to the dimer, bigger Cr clusters experience a reversal of their magnetic chirality at the vicinity of  magnetic skyrmions. Their energy landscape is found rich by hosting potentially a dual-behavior, i.e. multi-domains being pinning and repulsive. In Figure~\ref{fig:Cr_trimer_fcc_hcp}, we plot the binding-energy landscape of Cr trimers together with their magnetic state when the core of the skyrmion is at position (0,0). This demonstrates a striking effect: the phenotype of the skyrmion-defect interaction can be engineered utilizing the same Cr trimer via atom-by-atom manipulation of its shape. The overall behavior seems similar for both stacking sites with some distinct differences. Rotating the nanostructure by 180$^\circ$ from the pinning compact trimer-2 to trimer-1, leads to an opening of a repulsive domain in case of the fcc stacking, which grows in size when opening the trimer (corner-trimer) before ending up with a single repulsive domain for the linear trimer. The main difference between the fcc and hcp stacking occurs for Cr trimer-1. While the hcp trimer is strongly pinning, the fcc one induces a bi-domain with opposite interaction nature: pinning and repulsive. This behavior is triggered by the nature of the trimer's non-collinear spin-texture induced by the  magnetic frustration between the antiferromagnetic intra-defect and defect-substrate interactions as discussed in Supplementary Note 4.

Because of the strong intra-defect magnetic frustration, a  N\'eel state with a rotation angle of $120^\circ$ between the adatom moments is expected. Owing to the magnetic substrate, alteration of this state is observed depending on the shape and stacking of the defects. For instance, the hcp compact trimer (Fig.~\ref{fig:Cr_trimer_fcc_hcp}f) is in a N\'eel state with moments forming a spin-flop configuration resembling the one found for the dimer with moments lying almost in-plane and perpendicular to z-axis defined by the ferromagnetic magnetization direction. In strong contrast, the same trimer on the fcc stacking site has moments forging a ferrimagnetic-like non-collinear configuration (Fig.~\ref{fig:Cr_trimer_fcc_hcp}b), where the majority of atoms have their moments pointing almost antiferromagnetically to the substrate moments.

The simplified model addressed in Supplementary Note 4 gives a good account of the impurities magnetic texture and their stability, which is governed by the ratio of magnetic interactions $\text{J}_\text{ad-sub}/\text{J}_\text{ad-ad}$. The polar angles of two of the adatom moments is $164^\circ$, which is rather close to $136^\circ$ found from the model ($\cos{\theta} \approx -\frac{1}{2}  -\frac{3}{2}  \frac{\left<J_\text{ad-sub}\right>}{\left<J_\text{ad-ad}\right>}$) utilizing the averaged HEI listed in  Table~\ref{tab:jij}. At the skyrmion vicinity, the effective adatom-skyrmion exchange interaction decreases because of non-collinearity and therefore the polar angle decreases.  In the particular case of the distinct magnetic behavior of the fcc and hcp compact trimer-1, 
two mechanisms are at play: (i) the effective adatom-substrate magnetic interaction being reduced because  of the substrate's non-collinearity and (ii) the weaker ratio of magnetic interactions ($J_\text{ad-sub}/J_\text{ad-ad}$) for  hcp compared to the fcc stacking site, respectively $\sim 10\%$ and  $\sim 15\%$.

 The sequence of phenotypes observed by rotating and opening the trimers can further be explained with similar arguments. For instance, the immediate environment of trimer-2 and trimer-1 is different leading to a   decrease in the ratio of HEI, ($J_\text{ad-sub}/J_\text{ad-ad}$), for the former when compared to that of  the latter. By opening the trimer to form a corner-trimer, the spin-texture is less non-collinear and rather ferrimagnetic since the intra-defect magnetic frustration is reduced. The effective magnetic interaction felt by the surrounding substrate's atoms is anisotropic. Therefore, the stiffness of the skyrmion is enhanced close to the region seeing two of the parallel Cr magnetic moments. As  observed in Figure~\ref{fig:Cr_trimer_fcc_hcp}b-c-g,  repulsive areas emerge at the vicinity of the majority of magnetic moments pointing along a similar direction.

As expected from a previous study~\cite{Mavropoulos2010}, the antiferromagnetic interaction given in Table~\ref{tab:jij} tend to strongly decreases between the adatoms of the linear trimer, in comparison to that of the compact and corner trimers. The hybridization mechanisms between the electronic states of the adatoms weakens the HEI when the impurities form a chain. 
This favors a transition towards a collinear behavior with the three adatom moments aligned antiparallel to the substrate magnetization (see Figure~\ref{fig:Cr_trimer_fcc_hcp}). The antiferromagnetic interaction with the substrate is strong enough to impose such a collinear magnetic behavior. 
Here, the adatom-induced exchange contribution to the binding energy increases linearly with the number of adatoms, which induces an overall repulsive behavior. As expected, the linear  Cr-tetramer behaves similarly to the trimer line and acts as repulsive defect independently from the stacking site.

\subsection{Anatomy of defect-skyrmion interaction profiles: case of Fe-nanostructures.}
\begin{figure}
	\centering
	\includegraphics[width=\textwidth]{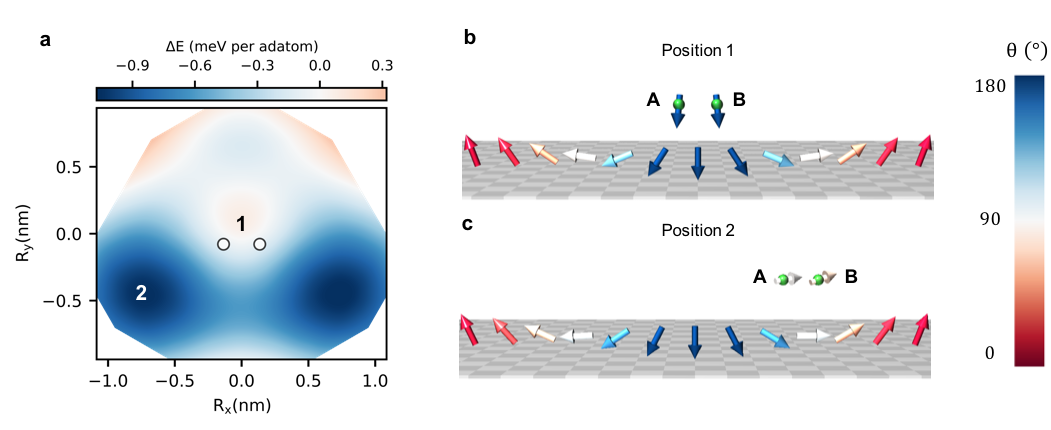}
	\caption{ \textbf{The skyrmion-defect energy profile induced by fcc Fe dimer-1.} \textbf{a} Multiple domains emerge owing to the subtle balance of the long-range magnetic interactions and the Zeeman energy. \textbf{b,c} Side views of the skyrmion-dimer complex, illustrating how the magnetic moments of the impurities adapt to the underlying skyrmion spin-texture. $\theta$ is the polar angle of the magnetic moments.} 
	\label{fig:FeDimer}
\end{figure}

Although being ferromagnetic in nature, Fe-clusters interacting with a single magnetic skyrmion  showcase 
a rich energy profile with a dual-behavior. 
 Similarly to Cr case, Fe dimer-2 and dimer-3 exhibit a similar behavior to the one of the adatom while the other nanostructures can display a complex energy profile. For several of the investigated Fe defects, the weakening of the substrate magnetic interaction, important for the pinning of skyrmions, is not strong enough to overcome the  repulsion from the exchange and the Zeeman contributions enabled by the Fe adatoms (see Supplementary Figure 5). While the Cr dimer-1 leads to a single pinning domain, two regions, pinning and repulsive, emerge from fcc Fe dimer-1  as shown in Figure~\ref{fig:FeDimer}a. 
The magnetic moments of the adatoms couple ferromagnetically to the Fe substrate, thus, when the skyrmion core is located underneath the defect (position 1) the non-collinearity of the substrate induces an opening angle of $\sim 26^\circ$ between the adatoms moments (see Figure~\ref{fig:FeDimer}b). In this particular magnetic configuration, a large unfavorable contribution  from the Zeeman interaction arises leading to the repulsion of the magnetic skyrmion since the adatoms moments are antiparallel to the applied magnetic field. Similarly to fcc Cr dimer-1, two off-centered pinning minima (see position 2 in Figure~\ref{fig:FeDimer}a) are obtained for Fe, which in this case is related to the Zeeman energy.  
The adatoms moments follow the underlying spin-texture of the skymion and point, therefore, almost inplane and perpendicular to the external magnetic field at position 2 (see Figure~\ref{fig:FeDimer}c). Hence, the mechanism leading to repulsion enabled by the Zeeman interaction is not anymore active. The same arguments can be used for all the more complex nanostructures made of Fe adatoms. When increasing the size of the nanostructures, single domains tend to develop. They are in general pinning for the fcc stacking site but switch to repulsion if the stacking is of hcp type. This is induced by the general weakening tendency of the adatom-substrate HEI when the stacking is of hcp instead of fcc type (see Table~\ref{tab:jij}).  While on the fcc stacking the adatom-substrate exchange interaction oscillate as function of the distance, on the hcp stacking the adatom couples ferromagnetic with both the first and the second nearest neighbour (see Supplementary Note 1). The latter provides a stronger magnetic exchange energy than for the fcc stacking sites, which stiffens the region under the defects and leads to the repulsion of the magnetic skyrmion.

\section*{Discussion}
\begin{figure}
	\centering
	\includegraphics[width=\textwidth]{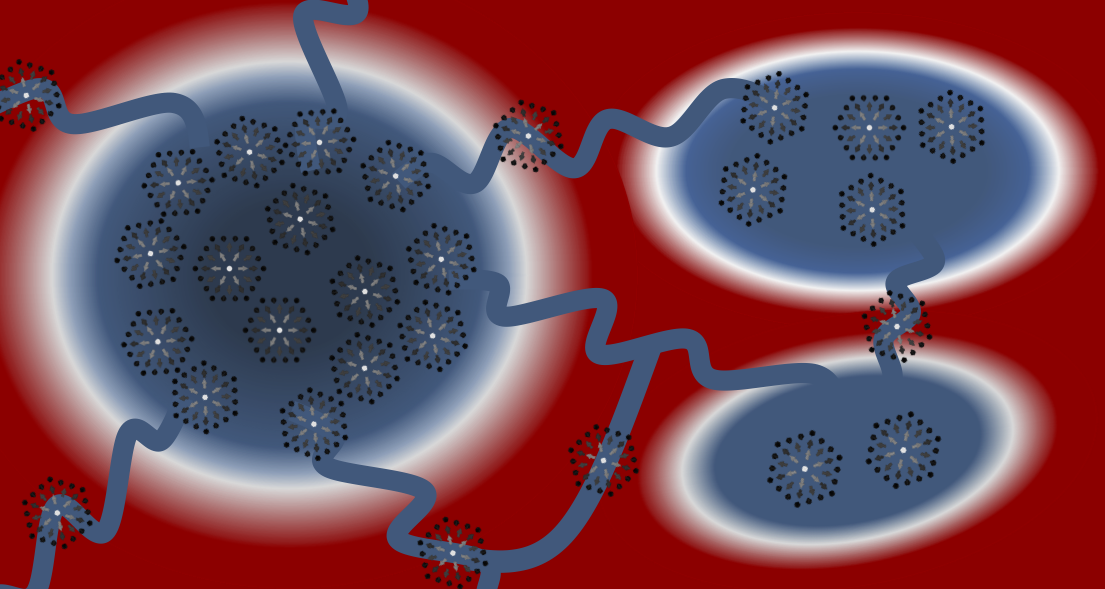}
	\caption{\textbf{A sketch of a skyrmion-based device patterned with multi-atom defects built atom-by-atom.} Design of the nanostructures shape and their proper positioning enable the emergence of multi-domains of opposite nature, pinning and repulsive. The same sets of defects can be utilized to create forbidden areas, eventually confining regions for reservoir computing and networks for efficient motion of skyrmions.}
	\label{fig:design_device}
\end{figure}

In this work, we have investigated the possibility of crafting skyrmion-defect interaction profiles by building-up multi-atomic defects atom-by-atom. Borrowing  concepts known in genetics, the mechanisms favoring pinning of repulsion of skyrmions define the interaction genotype carried by the building blocks of the defects, i.e. the adatoms.  The resulting skyrmion-defect interaction phenotype is found to be rich. It can be very different from the one of the single impurities and can potentially be engineered by controlling various parameters. Besides the obvious preponderant impact of the chemical nature of the impurities, we identified the shape, size and stacking of the defects to be essential knobs to tune the energy landscape of magnetic skyrmions governed by the interplay of impurity-substrate and the impurity-impurity magnetic interactions. When positioned in a nanostructure initially  repulsive, impurities develop, strikingly, a pinning or even a dual behavior by generating energy profiles hosting multi-domains.
 
Antiferromagnetic Cr nanostructures can host non-collinear spin-textures enabled by magnetic frustration due to the competition of intra-defect and defect-substrate magnetic interactions. The tendency for  non-collinearity of the defects is amplified at the vicinity of skyrmions. The smallest nanostructure is a dimer characterized by a spin-flop state, which facilitates the pinning of skyrmions. In fact, we conjecture a pinning interaction  phenotype for compact defects containing an even number of atoms since the total impurity-substrate magnetic interaction cancels out. The latter is finite for an odd number of impurities, which in general would promote ferrimagnetism and therefore repulsion. Fe nanostructures are ferromagnetic with rather sensitive magnetic interactions to the aforementioned knobs. As a result of the weakening of  impurity-substrate magnetic interactions, the dimers are found mostly pinning but developing small repulsive domains. Shifting the defects from fcc to hcp stacking switches them to a repulsive behavior.

We envisage atomic fabrication of nanostructures made of the same atoms to design complex energy landscapes for the benefit of skyrmion-based devices. Resting on already available  subnanoscale technologies,  racetracks could be patterned with defects in a Lego-fashion to obtain personalized energy landscapes. Out of a row of defects generating a dual pinning-repulsive behavior, one could divide the racetrack into regions with the opposite interaction nature and even manufacture inert areas where the interactions cancel each other. One may create confining regions for potential reservoir computing~\cite{Pinna2018} or build complex slalom-paths and networks to counteract properly the expected skyrmion Hall effect (see Figure~\ref{fig:design_device}). Our work promotes the exploration of man-controlled complex defects to exciting and useful constituents of  future nanotechnology   devices resting on non-collinear spin-textures.

\begin{methods}

\subsection{First-principles calculations.}
The first-principles calculations are based on density functional theory considering the local spin density approximation~\cite{Vosko1980}, as implemented in the full-potential relativistic Korringa-Kohn-Rostoker Green function (KKR) method~\cite{Papanikolaou2002,Bauer2013}. The latter allows us to embed the nanostructures  in a non-perturbed magnetic substrate by using a real-space embedding procedure. The adatoms are relaxed by 18\% similarly to what was considered in Ref.\cite{Fernandes2019}.  To extract the tensorial exchange coupling, one single interation was performed using the infinitesimal rotation method\cite{Ebert2009, Liechtenstein1987, Lounis2010} considering a k-mesh of $200 \times 200$ and an angular momentum cut-off at $l_\text{max} = 3$.
The Pd overlayer carries a sizable spin-moment of $\approx0.3 \mu_\text{B}$ induced by the Fe atoms, which is incorporated in the atomistic model via a renormalization scheme of the substrate's magnetic exchange interactions as described in Ref.\cite{PhysRevB.82.214409}.  The magnetic states are investigated via an atomistic spin-dynamics approach~\cite{skubic2008method,evans2014atomistic} extended to treat an embedding problem similar to the {\it ab initio} method. 

\end{methods}

\begin{addendum}
 \item We acknowledge discussions with Filipe Souza Mendes Guimar\~aes. This work is supported by the European Research Council (ERC) under the European Union's Horizon 2020 research and innovation programme (ERC-consolidator grant 681405 — DYNASORE). We gratefully acknowledge the computing time granted by JARA-HPC on the supercomputer JURECA at Forschungszentrum Jülich and by RWTH Aachen University.

\item[Author contributions] 
 S.L. initiated, designed and supervised the project. I.G.A.  performed the simulations with the help of I.L.F. and J.C. Intensive post-processing of the results was done by I.L.F. The atomistic spin-dynamics approach and code~\cite{skubic2008method,evans2014atomistic} were extended by J.C. to treat defects. All authors discussed the results. The manuscript was mainly written by I.L.F. and S.L. 
 
\item[Competing Interests] The authors declare that they have no competing financial interests. 

\item[Correspondence] Correspondence and requests for materials should be addressed to I.L.F. (email: i.lima.fernandes@fz-juelich.de) or to S.L. (email: s.lounis@fz-juelich.de).
\end{addendum}

%%
%% TABLES
%%

\begin{table}
\footnotesize
\centering
\caption{\textbf{Heisenberg magnetic exchange interactions for Cr and Fe nanostructures deposited on PdFe/Ir(111)}. The interactions (given in meV) are averaged among the nearest neighbors between the adatoms,  $\left<J_\text{ad-ad}\right>$, and between the adatoms and the nearest neighboring Fe atoms from the substrate,  $\left<J_\text{{ad-sub}}\right>$.  The superscript number indicates the number of atoms carrying the same moment.}
\begin{tabular}{c|c|c|c|c|c|c|c|c}
\hline
\hline
defect &site & $\left<J_\text{ad-ad}\right>_{\text{NN}}$ & $\left<J_\text{{ad-sub}}\right>$ & $\left<J_\text{sub}\right>$  &site & $\left<J_\text{ad-ad}\right>_{\text{NN}}$ & $\left<J_\text{{ad-sub}}\right>$ & $\left<J_\text{sub}\right>$  \\ \hline 
Cr adatom & fcc &- & -12.0 & 16.1 & hcp  & - & -11.6 & 15.3 \\ \hline
Cr dimer-1& fcc & -41.9 & -14.2 & 15.1 & hcp  & -43.3 & -14.3 &  16.4 \\ \hline
Cr dimer-2 & fcc & -0.36 & -12.3 & 15.7 & hcp & -1.3 & -12.0 & 15.8 \\ \hline
Cr dimer-3 & fcc & 0.16 & -12.0 & 16.1 & hcp & 0.05 & -11.6 & 15.4  \\ \hline
Cr compact trimer-1 & fcc & -62.7 & -9.3 & 14.9 & hcp & -55.9 & -5.6 & 15.9 \\ \hline
Cr compact trimer-2 & fcc & -54.7 & -6.9 & 15.5 & hcp & -59.4 & -7.9 & 16.1 \\ \hline
Cr corner trimer & fcc & -54.5 & -3.8$^1$, -10.1$^2$ & 15.1 & hcp & -38.4 & -12.1$^1$, -9.7$^2$  & 8.6  \\ \hline
Cr linear trimer & fcc & -24.3 & -19.3$^1$, -15.8$^2$  & 14.5 & hcp & -28.7 & -20.2$^1$, -14.5$^2$ & 15.9 \\ \hline
Cr linear tetramer & fcc & -15.7$^2$, -23.8$^2$ & -19.7$^2$,-15.6$^2$ & 13.3 & hcp & -18.2$^2$, -27.7$^2$ & -19.5$^2$, -14.3$^2$  & 13.5 \\ \hline \hline

Fe adatom & fcc & - & 8.4 & 15.6 & hcp & - & 7.3 & 13.9 \\ \hline
Fe dimer-1 & fcc & 36.8 & 7.4 & 15.2 & hcp & 33.5 & 10.3 &  16.0 \\ \hline
Fe dimer-2 & fcc & 0.2 & 8.4 & 15.0 & hcp & 0.2 & 7.5 & 15.4 \\ \hline
Fe dimer-3 & fcc & -0.3 & 8.3 & 16.0 & hcp & 33.5 & 10.3 &  16.0 \\ \hline
Fe compact trimer-1 & fcc & 36.3 & 6.2 & 15.1 & hcp & 37.1 & 7.9 & 15.4 \\ \hline
Fe compact trimer-2 & fcc & 33.7 & 5.8 & 14.6 & hcp & 32.8 & 13.1 & 15.9 \\ \hline
Fe corner trimer & fcc & 31.5 & 5.9$^1$, 7.6$^2$ & 14.8 & hcp & 27.7 & 11.1$^1$, 9.8$^2$ & 17.2 \\ \hline
Fe linear trimer & fcc & 26.9 & 6.1$^1$, 7.6$^2$ & 14.7 & hcp & 21.4 & 11.8$^1$, 9.6$^2$& 15.6 \\ \hline 
Fe linear tetramer & fcc & 18.1$^2$, 28.4$^2$ & 5.8$^2$, 7.5$^2$ & 14.5 & hcp & 10.4$^2$, 23.8$^2$ & 10.8$^2$, 9.7$^2$ & 15.3  \\ \hline \hline
\end{tabular}
\label{tab:jij}
\end{table}

\end{document}